\documentclass[twocolumn,showpacs,aps,pra,amsmath,amssymb]{revtex4}%
\usepackage{graphicx}
\usepackage{amsfonts}
\usepackage{amsmath}
\usepackage{amssymb}%
\newtheorem{theorem}{Theorem}
\newtheorem{corollary}[theorem]{Corollary}

\newtheorem{lemma}[theorem]{Lemma}
\newtheorem{proposition}[theorem]{Proposition}
\newenvironment{proof}[1][Proof]{\textbf{#1.} }{\ \rule{0.5em}{0.5em}}
\begin{document}
\title{Improved Simulation of Stabilizer Circuits}
\author{Scott Aaronson}
\email{aaronson@csail.mit.edu}
\affiliation{MIT}
\author{Daniel Gottesman}
\email{dgottesman@perimeterinstitute.ca}
\affiliation{Perimeter Institute}

\begin{abstract}
The Gottesman-Knill theorem says that a stabilizer circuit---that is, a
quantum circuit consisting solely of CNOT, Hadamard, and phase gates---can be
simulated efficiently on a classical computer. \ This paper improves that
theorem in several directions. \ First, by removing the need for Gaussian
elimination, we make the simulation algorithm much faster at the cost of a
factor-$2$ increase in the number of bits needed to represent a state. We have
implemented the improved algorithm in a freely-available program called CHP
(CNOT-Hadamard-Phase), which can handle thousands of qubits easily. \ Second,
we show that the problem of simulating stabilizer circuits is complete for the
classical complexity class $\mathsf{\oplus L}$, which means that stabilizer
circuits are probably not even universal for \textit{classical} computation.
\ Third, we give efficient algorithms for computing the inner product between
two stabilizer states, putting any $n$-qubit stabilizer circuit into a
\textquotedblleft canonical form\textquotedblright\ that requires at most
$O\left(  n^{2}/\log n\right)  $ gates, and other useful tasks. \ Fourth, we
extend our simulation algorithm to circuits acting on mixed states, circuits
containing a limited number of non-stabilizer gates, and circuits acting on
general tensor-product initial states but containing only a limited number of measurements.

\end{abstract}

\maketitle


\section{Introduction\label{INTRO}}

Among the many difficulties that quantum computer architects face, one of them
is almost intrinsic to the task at hand: how do you design and debug circuits
that you can't even simulate efficiently with existing tools? \ Obviously, if
a quantum computer output the factors of a $3000$-digit number, then you
wouldn't need to simulate it to verify its correctness, since multiplying is
easier than factoring. \ But what if the quantum computer \textit{didn't}
work? \ Ordinarily architects might debug a computer by adding test
conditions, monitoring registers, halting at intermediate steps, and so on.
\ But for a quantum computer, all of these standard techniques would probably
entail measurements that destroy coherence. \ Besides, it would be nice to
design and debug a quantum computer using classical CAD tools, \textit{before}
trying to implement it!

Quantum architecture is one motivation for studying classical algorithms to
simulate and manipulate quantum circuits, but it is not the only motivation.
\ Chemists and physicists have long needed to simulate quantum systems, and
they have not had the patience to wait for a quantum computer to be built.
\ Instead, they have developed limited techniques such as Quantum Monte-Carlo
(QMC) \cite{suzuki} for computing properties of certain ground states. \ More
recently, several general-purpose quantum computer simulators have appeared,
including Oemer's quantum programming language QCL \cite{oemer}, the QuIDD
(Quantum Information Decision Diagrams) package of Viamontes et al.
\cite{vmh,vrmh}, and the parallel quantum computer simulator of Obenland and
Despain \cite{od}. \ The drawback of such simulators, of course, is that their
running time grows exponentially in the number of qubits. \ This is true not
only in the worst case but in practice. \ For example, even though it uses a
variant of binary decision diagrams to avoid storing an entire amplitude
vector for some states, Viamontes et al. \cite{vmh}\ report that\ the QuIDD
package took more than 22 hours to simulate Grover's algorithm on 40 qubits.
\ With a general-purpose package, then, simulating hundreds or thousands of
qubits is out of the question.

A different direction of research has sought to find nontrivial classes of
quantum circuits that \textit{can} be simulated efficiently on a classical
computer. \ For example, Vidal \cite{vidal}\ showed that, so long as a quantum
computer's state at every time step has polynomially-bounded entanglement
under a measure related to Schmidt rank, the computer can be simulated
classically in polynomial time. \ Notably, in a follow-up paper \cite{vidal2}%
,\ Vidal actually implemented his algorithm and used it to simulate
$1$-dimensional quantum spin chains consisting of hundreds of spins.\ \ A
second example is a result of Valiant \cite{valiant}, which reduces the
problem of simulating a restricted class of quantum computers to that of
computing the Pfaffian of a matrix. \ The latter is known to be solvable in
classical polynomial time. \ Terhal and DiVincenzo \cite{td} have shown that Valiant's class corresponds to
a model of noninteracting fermions.

There is one class of quantum circuits that is known to be simulable in
classical polynomial time, that does not impose any limit on entanglement, and
that arises naturally in several applications. \ This is the class of
\textit{stabilizer circuits}\ introduced to analyze quantum error-correcting
codes \cite{bdsw,crss,gottesman,gottesman2}. \ A stabilizer circuit is simply
a quantum circuit in which every gate is a controlled-NOT, Hadamard, phase, or
$1$-qubit measurement gate. \ We call a stabilizer circuit \textit{unitary} if
it does not contain measurement gates. \ Unitary stabilizer circuits are also
known as Clifford group circuits.%

\begin{figure}
[ptb]
\begin{center}
\includegraphics[
trim=0.483056in 3.658617in 0.682532in 0.000000in,
height=1.472in,
width=3.1673in
]%
{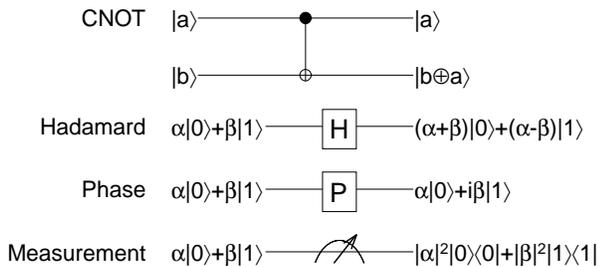}%
\caption{The four types of gate allowed in the stabilizer formalism}%
\label{gates}%
\end{center}
\end{figure}

Stabilizer circuits can be used to perform the encoding and
decoding steps for a quantum error-correcting code, and they play an important
role in fault-tolerant circuits. \ However, the
\textit{stabilizer formalism} used to describe these circuits has many other
applications. \ This formalism is rich enough to encompass most of
the \textquotedblleft paradoxes\textquotedblright\ of quantum mechanics,
including the GHZ (Greenberger-Horne-Zeilinger) experiment \cite{ghz}, dense
quantum coding \cite{bw}, and quantum teleportation \cite{bbcjpw}. \ On the
other hand, it is not \textit{so} rich as to preclude efficient simulation by
a classical computer. \ That conclusion, sometimes known as the
\textit{Gottesman-Knill theorem}, is the starting point for the contributions
of this paper.

Our results are as follows. \ In Section \ref{SIM}\ we give a new
\textit{tableau algorithm} for simulating stabilizer circuits that is faster
than the algorithm directly implied by the Gottesman-Knill theorem. \ By
removing the need for Gaussian elimination, this algorithm enables
measurements to be simulated in $O\left(  n^{2}\right)  $\ steps instead of
$O\left(  n^{3}\right)  $\ (where $n$ is the number of qubits), at a cost of a
factor-$2$ increase in the number of bits needed to represent a quantum state.

Section \ref{IMPLEM} describes CHP, a high-performance stabilizer circuit
simulator that implements our tableau algorithm.\ \ We present the results of
an experiment designed to test how CHP's performance is affected by properties
of the stabilizer circuit being simulated. \ CHP has already found application
in simulations of quantum fault-tolerance circuits \cite{cross}.

Section \ref{PARITYL}\ proves that the problem of simulating stabilizer
circuits is complete for the classical complexity class $\mathsf{\oplus L}$.
\ Informally, this means that any stabilizer circuit can be simulated using
CNOT gates alone; the availability of Hadamard and phase gates provides at
most a polynomial advantage. \ This result removes some of the mystery about
the Gottesman-Knill theorem by showing that stabilizer circuits are unlikely
to be capable even of universal \textit{classical} computation.

In Section \ref{CANONICAL}\ we prove a \textit{canonical form theorem} that we
expect will have many applications to the study of stabilizer circuits. \ The
theorem says that given any stabilizer circuit, there exists an equivalent
stabilizer circuit that applies a round of Hadamard gates, followed by a round
of phase gates, followed by a round of CNOT gates, and so on in the sequence
H-C-P-C-P-C-H-P-C-P-C (where H, C, P stand for Hadamard, CNOT, Phase
respectively). \ One immediate corollary, building on a result by Patel,
Markov, and Hayes \cite{pmh} and improving one by Dehaene and De Moor
\cite{dm}, is that any stabilizer circuit on $n$ qubits has an equivalent
circuit with only $O\left(  n^{2}/\log n\right)  $\ gates.

Finally, Section \ref{BEYOND} extends our simulation algorithm to situations
beyond the usual one considered in the Gottesman-Knill theorem. \ For example,
we show how to handle mixed states, \textit{without} keeping track of pure
states from which the mixed states are obtainable by discarding qubits. \ We
also show how to simulate circuits involving a small number of non-stabilizer
gates; or involving arbitrary tensor-product initial states, but only a small
number of measurements. \ Both of these latter two simulations take time that
is polynomial in the number of qubits, but exponential in the number of
non-stabilizer gates or measurements. \ Presumably this exponential dependence
is necessary, since otherwise we could simulate arbitrary quantum computations
in classical subexponential time.

We conclude in Section \ref{OPEN}\ with some directions for further research.

\section{Preliminaries\label{PRELIM}}

We assume familiarity with quantum computing. \ This section provides a crash
course on the stabilizer formalism, confining attention to those aspects we
will need. \ See Section 10.5.1 of Nielsen and Chuang \cite{nc} for more details.

Throughout this paper we will use the following four Pauli matrices:%
\[%
\begin{array}
[c]{ll}%
I=\left(
\begin{array}
[c]{cc}%
1 & 0\\
0 & 1
\end{array}
\right)  & X=\left(
\begin{array}
[c]{cc}%
0 & 1\\
1 & 0
\end{array}
\right) \\
Y=\left(
\begin{array}
[c]{cc}%
0 & -i\\
i & 0
\end{array}
\right)  & Z=\left(
\begin{array}
[c]{cc}%
1 & 0\\
0 & -1
\end{array}
\right)
\end{array}
\]
These matrices satisfy the following identities:%
\begin{align*}
&  X^{2}=Y^{2}=Z^{2}=I\\
&
\begin{array}
[c]{ccc}%
XY=iZ & YZ=iX & ZX=iY\\
YX=-iZ & ZY=-iX & XZ=-iY
\end{array}
\end{align*}
In particular, every two Pauli matrices either commute or anticommute. \ The
rule for whether to include a minus sign is the same as that for quaternions,
if we replace $\left(  I,X,Y,Z\right)  $ by $\left(  1,i,j,k\right)  $.

We define the group $\mathcal{P}_{n}$\ of $n$-qubit\textit{ Pauli operators}
to consist of all tensor products of $n$ Pauli\ matrices, together with a
multiplicative factor of $\pm1$ or $\pm i$\ (so the total number of operators
is $\left|  \mathcal{P}_{n}\right|  =4^{n+1}$). \ We omit tensor product signs
for brevity; thus $-YZZI$\ should be read $-Y\otimes Z\otimes Z\otimes I$ (we
will use $+$\ to represent the Pauli group operation). \ Given two Pauli
operators $P=i^{k}P_{1}\cdots P_{n}$\ and $Q=i^{l}Q_{1}\cdots Q_{n}$, it is
immediate that $P$ commutes with $Q$ if and only if the number of indices
$j\in\left\{  1,\ldots,n\right\}  $\ such that $P_{j}$\ anticommutes with
$Q_{j}$ is even; otherwise $P$ anticommutes with $Q$. \ Also, for all
$P\in\mathcal{P}_{n}$, if $P$ has a phase of $\pm1$\ then $P+P=I\cdots I$,
whereas if $P$ has a phase of $\pm i$\ then $P+P=-I\cdots I$.

Given a pure quantum state $\left|  \psi\right\rangle $,\ we say a unitary
matrix $U$ \textit{stabilizes} $\left|  \psi\right\rangle $\ if $\left|
\psi\right\rangle $\ is an eigenvector of $U$\ with eigenvalue $1$, or
equivalently if$\ U\left|  \psi\right\rangle =\left|  \psi\right\rangle $
where we do not ignore global phase. \ To illustrate, the following table
lists the Pauli matrices and their opposites, together with the unique
$1$-qubit states that they stabilize:%
\[%
\begin{array}
[c]{lllll}%
X: & \left|  0\right\rangle +\left|  1\right\rangle  & \,\,\, & -X: & \left|
0\right\rangle -\left|  1\right\rangle \\
Y: & \left|  0\right\rangle +i\left|  1\right\rangle  & \,\,\, & -Y: & \left|
0\right\rangle -i\left|  1\right\rangle \\
Z: & \left|  0\right\rangle  & \,\,\, & -Z: & \left|  1\right\rangle
\end{array}
\]
The identity matrix $I$ stabilizes all states, whereas $-I$\ stabilizes no states.

The key idea of the stabilizer formalism is to represent a quantum state
$\left\vert \psi\right\rangle $, not by a vector of amplitudes, but by a
\textit{stabilizer group}, consisting of unitary matrices that stabilize
$\left\vert \psi\right\rangle $. \ Notice that if $U$ and $V$ both stabilize
$\left\vert \psi\right\rangle $ then so do $UV$ and $U^{-1}$, and thus the set
$\operatorname*{Stab}\left(  \left\vert \psi\right\rangle \right)  $\ of
stabilizers of $\left\vert \psi\right\rangle $\ is a group. \ Also, it is not
hard to show that if $\left\vert \psi\right\rangle \neq\left\vert
\varphi\right\rangle $\ then $\operatorname*{Stab}\left(  \left\vert
\psi\right\rangle \right)  \neq\operatorname*{Stab}\left(  \left\vert
\varphi\right\rangle \right)  $. \ But why does this strange representation
buy us anything?\ \ To write down generators for $\operatorname*{Stab}\left(
\left\vert \psi\right\rangle \right)  $\ (even approximately) still takes
exponentially many bits in general by an information-theoretic argument.
\ Indeed stabilizers seem \textit{worse} than amplitude vectors, since they
require about $2^{2n}$\ parameters to specify instead of about $2^{n}$!

Remarkably, though, a large and interesting class of quantum states can be
specified uniquely by much smaller stabilizer groups---specifically, the
intersection of $\operatorname*{Stab}\left(  \left|  \psi\right\rangle
\right)  $ with the Pauli group \cite{gottesman,gottesman2,crss}. \ This class
of states, which arises in quantum error correction and many other settings,
is characterized by the following theorem.

\begin{theorem}
\label{gk}Given an $n$-qubit state $\left|  \psi\right\rangle $, the following
are equivalent: \vspace{-0.05in}

\begin{enumerate}
\item[(i)] $\left|  \psi\right\rangle $ can be obtained from $\left|
0\right\rangle ^{\otimes n}$\ by CNOT, Hadamard, and phase gates only.
\vspace{-0.05in}

\item[(ii)] $\left|  \psi\right\rangle $ can be obtained from $\left|
0\right\rangle ^{\otimes n}$\ by CNOT, Hadamard, phase, and measurement gates
only. \vspace{-0.05in}

\item[(iii)] $\left|  \psi\right\rangle $ is stabilized by exactly $2^{n}%
$\ Pauli operators. \vspace{-0.05in}

\item[(iv)] $\left|  \psi\right\rangle $ is uniquely determined by $S\left(
\left|  \psi\right\rangle \right)  =\operatorname*{Stab}\left(  \left|
\psi\right\rangle \right)  \cap\mathcal{P}_{n}$, or the group of Pauli
operators that stabilize $\left|  \psi\right\rangle $.
\end{enumerate}
\end{theorem}

Because of Theorem \ref{gk}, we call any circuit consisting entirely of CNOT,
Hadamard, phase, and measurement gates a \textit{stabilizer circuit}, and any
state obtainable by applying a stabilizer circuit to $\left\vert
0\right\rangle ^{\otimes n}$\ a \textit{stabilizer state}. \ As a warmup to
our later results, the following proposition counts the number of stabilizer states.

\begin{proposition}
\label{count}Let $N$ be the number of pure stabilizer states on $n$ qubits.
\ Then%
\[
N=2^{n}\prod_{k=0}^{n-1}\left(  2^{n-k}+1\right)  =2^{\left(  1/2+o\left(
1\right)  \right)  n^{2}}.
\]

\end{proposition}

\begin{proof}
We have $N=G/A$, where $G$ is the total number of generating sets and $A$ is
the number of equivalent generating sets for a given stabilizer $S$. \ To find
$G$, note that there are $4^{n}-1$ choices for the first generator $M_{1}$
(ignoring overall sign), because it can be anything but the identity. \ The
second generator must commute with $M_{1}$ and cannot be $I$ or $M_{1}$, so
there are $4^{n}/2-2$ choices for $M_{2}$. \ Similarly, $M_{3}$ must commute
with $M_{1}$ and $M_{2}$, but cannot be in the group generated by them, so
there are $4^{n}/4-4$ choices for it, and so on. \ Hence, including overall
signs,%
\[
G=2^{n}\prod_{k=0}^{n-1}\left(  \frac{4^{n}}{2^{k}}-2^{k}\right)  =2^{n\left(
n+1\right)  /2}\prod_{k=0}^{n-1}\left(  4^{n-k}-1\right)  .
\]
Similarly, to find $A$, note that given $S$, there are $2^{n}-1$ choices for
$M_{1}$, $2^{n}-2$ choices for $M_{2}$, $2^{n}-4$ choices for $M_{3}$, and so
on. \ Thus
\[
A=\prod_{k=0}^{n-1}\left(  2^{n}-2^{k}\right)  =2^{n\left(  n-1\right)
/2}\prod_{k=0}^{n-1}\left(  2^{n-k}-1\right)  .
\]
Therefore
\[
N=\frac{G}{A}=2^{n}\prod_{k=0}^{n-1}\left(  \frac{4^{n-k}-1}{2^{n-k}%
-1}\right)  =2^{n}\prod_{k=0}^{n-1}\left(  2^{n-k}+1\right)  .
\]

\end{proof}

\section{Efficient Simulation of Stabilizer Circuits\label{SIM}}

Theorem \ref{gk}\ immediately suggests a way to simulate stabilizer circuits
efficiently on a classical computer. \ A well-known fact from group theory
says that any finite group $G$ has a generating set of size at most $\log
_{2}\left\vert G\right\vert $. \ So if $\left\vert \psi\right\rangle $\ is a
stabilizer state on $n$ qubits, then the group\ $S\left(  \left\vert
\psi\right\rangle \right)  $\ of Pauli operators that stabilize $\left\vert
\psi\right\rangle $ has a generating set of size $n=\log_{2}2^{n}$. \ Each
generator takes $2n+1$\ bits to specify: $2$\ bits for each of the $n$ Pauli
matrices, and $1$ bit for the phase \footnote{If $P\in S\left(  \left\vert
\psi\right\rangle \right)  $, then $P$ can only have a phase of $\pm1$, not
$\pm i$: for in the latter case $P^{2}=-I\cdots I$\ would be in $S\left(
\left\vert \psi\right\rangle \right)  $,\ but we saw that $-I$\ does not
stabilize anything.}. \ So the total number of bits needed to specify
$\left\vert \psi\right\rangle $\ is $n\left(  2n+1\right)  $. \ What Gottesman
and Knill showed, furthermore, is that these bits can be \textit{updated} in
polynomial time after a CNOT, Hadamard, phase, or measurement gate is applied
to $\left\vert \psi\right\rangle $. \ The updates corresponding to unitary
gates are very efficient, requiring only $O\left(  n\right)  $\ time for each gate.

However, the updates corresponding to measurements are not so efficient. \ We
can decide in $O\left(  n\right)  $\ time whether a measurement of qubit $a$
will yield a deterministic or random outcome. \ If the outcome is random, then
updating the state after the measurement takes $O\left(  n^{2}\right)
$\ time, but if the outcome is deterministic, then deciding whether the
outcome is $\left\vert 0\right\rangle $ or $\left\vert 1\right\rangle $ seems
to require inverting an $n\times n$\ matrix, which takes $O\left(
n^{2.376}\right)  $\ time in theory \cite{cw}\ but order $n^{3}$\ time in
practice. \ What that $n^{3}$\ complexity means is that simulations of, say,
$2000$-qubit\ systems would already be prohibitive on a desktop PC, given that
measurements are frequent.

This section describes a new simulation algorithm, by which both deterministic
and random measurements can be performed in $O\left(  n^{2}\right)  $ time.
\ The cost is a factor-$2$ increase in the number of bits needed to specify a
state. \ For in addition to the $n$ stabilizer generators, we now store
$n$\ \textquotedblleft destabilizer\textquotedblright\ generators, which are
Pauli operators that together with the stabilizer generators generate the full
Pauli group $\mathcal{P}_{n}$. \ So the number of bits needed is $2n\left(
2n+1\right)  \approx4n^{2}$.

The algorithm represents a state by a \textit{tableau}\ consisting of binary
variables $x_{ij},z_{ij}$ for all $i\in\left\{  1,\ldots,2n\right\}  $,
$j\in\left\{  1,\ldots,n\right\}  $, and $r_{i}$\ for all $i\in\left\{
1,\ldots,2n\right\}  $ \footnote{Dehaene and De Moor \cite{dm}\ came up with
something like this tableau representation independently, though they did not
use it to simulate measurements in $O\left(  n^{2}\right)  $\ time.}:%
\[
\left(
\begin{tabular}
[c]{ccc|ccc|c}%
$x_{11}$ & $\cdots$ & $x_{1n}$ & $z_{11}$ & $\cdots$ & $z_{1n}$ & $r_{1}$\\
$\vdots$ & $\ddots$ & $\vdots$ & $\vdots$ & $\ddots$ & $\vdots$ & $\vdots$\\
$x_{n1}$ & $\cdots$ & $x_{nn}$ & $z_{n1}$ & $\cdots$ & $z_{nn}$ & $r_{n}%
$\\\hline
$x_{\left(  n+1\right)  1}$ & $\cdots$ & $x_{\left(  n+1\right)  n}$ &
$z_{\left(  n+1\right)  1}$ & $\cdots$ & $z_{\left(  n+1\right)  n}$ &
$r_{n+1}$\\
$\vdots$ & $\ddots$ & $\vdots$ & $\vdots$ & $\ddots$ & $\vdots$ & $\vdots$\\
$x_{\left(  2n\right)  1}$ & $\cdots$ & $x_{\left(  2n\right)  n}$ &
$z_{\left(  2n\right)  1}$ & $\cdots$ & $z_{\left(  2n\right)  n}$ & $r_{2n}$%
\end{tabular}
\ \ \ \ \right)
\]
Rows $1$ to $n$ of the tableau represent the destabilizer generators
$R_{1},\ldots,R_{n}$, and rows $n+1$\ to $2n$ represent the stabilizer
generators $R_{n+1},\ldots,R_{2n}$. \ If $R_{i}=\pm P_{1}\cdots P_{n}$, then
bits $x_{ij},z_{ij}$\ determine the\ $j^{th}$\ Pauli matrix $P_{j}$: $00$
means $I$, $01$ means $X$, $11$ means $Y$, and $10$ means $Z$. \ Finally,
$r_{i}$\ is $1$ if $R_{i}$\ has negative phase and $0$ if $r_{i}$\ has
positive phase. \ As an example, the $2$-qubit state $\left\vert
00\right\rangle $\ is stabilized by the Pauli operators $+ZI$\ and $+IZ$, so a
possible tableau for $\left\vert 00\right\rangle $\ is%
\[
\left(
\begin{tabular}
[c]{cc|cc|c}%
$~1~$ & $~0~$ & $~0~$ & $~0~$ & $~0~$\\
$~0~$ & $~1~$ & $~0~$ & $~0~$ & $~0~$\\\hline
$~0~$ & $~0~$ & $~1~$ & $~0~$ & $~0~$\\
$~0~$ & $~0~$ & $~0~$ & $~1~$ & $~0~$%
\end{tabular}
\right)
\]
Indeed, we will take the obvious generalization of the above \textquotedblleft
identity matrix\textquotedblright\ to be the standard initial tableau.

The algorithm uses a subroutine called $\operatorname*{rowsum}\left(
h,i\right)  $, which sets generator $h$ equal to $i+h$. \ Its purpose is to
keep track, in particular, of the phase bit $r_{h}$, including all the factors
of $i$ that appear when multiplying Pauli matrices. The subroutine is
implemented as follows.\medskip

\noindent\textbf{rowsum}$\left(  h,i\right)  $: Let $g\left(  x_{1}%
,z_{1},x_{2},z_{2}\right)  $\ be a function that takes $4$ bits as input, and
that returns the exponent to which $i$ is raised (either $0$, $1$, or $-1$)
when the Pauli matrices represented by $x_{1}z_{1}$\ and $x_{2}z_{2}$\ are
multiplied. \ More explicitly, if $x_{1}=z_{1}=0$\ then $g=0$; if $x_{1}%
=z_{1}=1$\ then $g=z_{2}-x_{2}$; if $x_{1}=1$\ and $z_{1}=0$\ then
$g=z_{2}\left(  2x_{2}-1\right)  $; and if $x_{1}=0$\ and $z_{1}=1$\ then
$g=x_{2}\left(  1-2z_{2}\right)  $. \ Then set $r_{h}:=0$\ if%
\[
2r_{h}+2r_{i}+\sum_{j=1}^{n}g\left(  x_{ij},z_{ij},x_{hj},z_{hj}\right)
\equiv0\left(  \operatorname{mod}4\right)  ,
\]
and set $r_{h}:=1$\ if the sum is congruent to $2$\ mod $4$ (it will never be
congruent to $1$ or $3$). \ Next, for all $j\in\left\{  1,\ldots,n\right\}
$,\ set $x_{hj}:=x_{ij}\oplus x_{hj}$ and set $z_{hj}:=z_{ij}\oplus z_{hj}$
(here and throughout, $\oplus$\ denotes exclusive-OR).\medskip

We now give the algorithm. \ It will be convenient to add an additional
$\left(  2n+1\right)  ^{st}$\ row for \textquotedblleft scratch
space.\textquotedblright\ \ The initial state $\left\vert 0\right\rangle
^{\otimes n}$\ has $r_{i}=0$\ for all $i\in\left\{  1,\ldots,2n+1\right\}  $,
and $x_{ij}=\delta_{ij}$\ and $z_{ij}=\delta_{\left(  i-n\right)  j}$\ for all
$i\in\left\{  1,\ldots,2n+1\right\}  $ and\ $j\in\left\{  1,\ldots,n\right\}
$, where $\delta_{ij}$ is $1$\ if $i=j$\ and $0$ otherwise. \ The algorithm
proceeds through the gates in order; for each one it does one of the following
depending on the gate type.\medskip

\noindent\textbf{CNOT from control }$a$\textbf{ to target }$b$\textbf{.} \ For
all $i\in\left\{  1,\ldots,2n\right\}  $, set $r_{i}:=r_{i}\oplus x_{ia}%
z_{ib}\left(  x_{ib}\oplus z_{ia}\oplus1\right)  $, $x_{ib}:=x_{ib}\oplus
x_{ia}$, and $z_{ia}:=z_{ia}\oplus z_{ib}$.\medskip

\noindent\textbf{Hadamard on qubit }$a$\textbf{.} \ For all $i\in\left\{
1,\ldots,2n\right\}  $, set $r_{i}:=r_{i}\oplus x_{ia}z_{ia}$\ and swap
$x_{ia}$\ with $z_{ia}$.\medskip

\noindent\textbf{Phase on qubit }$a$\textbf{.} \ For all $i\in\left\{
1,\ldots,2n\right\}  $, set $r_{i}:=r_{i}\oplus x_{ia}z_{ia}$ and then set
$z_{ia}:=z_{ia}\oplus x_{ia}$.\medskip

\noindent\textbf{Measurement of qubit }$a$\textbf{ in standard basis.} \ First
check whether there exists a $p\in\left\{  n+1,\ldots,2n\right\}  $\ such that
$x_{pa}=1$.

\noindent\textbf{Case I:} Such a $p$ exists (if more than one exists, then let
$p$ be the smallest). \ In this case the measurement outcome is random, so the
state needs to be updated. \ This is done as follows. \ First call
$\operatorname*{rowsum}\left(  i,p\right)  $\ for all $i\in\left\{
1,\ldots,2n\right\}  $\ such that $i\neq p$\ and $x_{ia}=1$.\ \ Second, set
entire the $\left(  p-n\right)  ^{th}$\ row equal to the $p^{th}$\ row.
\ Third, set the $p^{th}$\ row to be identically $0$, except\ that $r_{p}$\ is
$0$ or $1$ with equal probability, and $z_{pa}=1$. \ Finally, return $r_{p}%
$\ as the measurement outcome.

\noindent\textbf{Case II:} Such an $p$ does not exist. \ In this case the
outcome is determinate, so measuring the state will not change it; the only
task is to determine whether $0$ or $1$ is observed. \ This is done as
follows. \ First set the $\left(  2n+1\right)  ^{st}$\ row to be identically
$0$. \ Second, call $\operatorname*{rowsum}\left(  2n+1,i+n\right)  $\ for all
$i\in\left\{  1,\ldots,n\right\}  $\ such that $x_{ia}=1$. \ Finally return
$r_{2n+1}$\ as the measurement outcome.\medskip

Once we interpret the $x_{ij}$, $z_{ij}$, and $r_{i}$\ bits for $i\geq n+1$ as
representing generators of $S\left(  \left|  \psi\right\rangle \right)  $, and
$\operatorname*{rowsum}$\ as representing the group operation in
$\mathcal{P}_{n}$, the correctness of the CNOT, Hadamard, phase, and random
measurement procedures follows immediately from previous analyses by Gottesman
\cite{gottesman2}. \ It remains only to explain why the determinate
measurement procedure is correct. \ Observe that $R_{h}$\ commutes with
$R_{i}$\ if the \textit{symplectic inner product}%
\[
R_{h}\cdot R_{i}=x_{h1}z_{i1}\oplus\cdots\oplus x_{hn}z_{in}\oplus
x_{i1}z_{h1}\oplus\cdots\oplus x_{in}z_{hn}%
\]
equals $0$, and anticommutes with $R_{i}$\ if $R_{h}\cdot R_{i}=1$. \ Using
that fact it is not hard to show the following.

\begin{proposition}
\label{invariant}The following are invariants of the tableau algorithm:
\vspace{-0.05in}

\begin{itemize}
\item[(i)] \noindent$R_{n+1},\ldots,R_{2n}$ generate $S\left(  \left|
\psi\right\rangle \right)  $, and $R_{1},\ldots,R_{2n}$\ generate
$\mathcal{P}_{n}$. \vspace{-0.05in}

\item[(ii)] \noindent$R_{1},\ldots,R_{n}$ commute. \vspace{-0.05in}

\item[(iii)] \noindent For all $h\in\left\{  1,\ldots,n\right\}  $, $R_{h}%
$\ anticommutes with $R_{h+n}$. \vspace{-0.05in}

\item[(iv)] \noindent For all $i,h\in\left\{  1,\ldots,n\right\}  $ such that
$i\neq h$, $R_{i}$\ commutes with $R_{h+n}$.
\end{itemize}
\end{proposition}

Now suppose that a measurement of qubit $a$ yields a determinate outcome.
\ Then the $Z_{a}$\ operator must commute with all elements of the stabilizer,
so%
\[
\sum_{h=1}^{n}c_{h}R_{h+n}=\pm Z_{a}%
\]
for a unique choice of $c_{1},\ldots,c_{n}\in\left\{  0,1\right\}  $. \ Our
goal is to determine the $c_{h}$'s, since then by summing the appropriate
$R_{h+n}$'s we can learn whether the phase representing the outcome is
positive or negative. \ Notice that for all $i\in\left\{  1,\ldots,n\right\}
$,%
\[
c_{i}\equiv\sum_{h=1}^{n}c_{h}\left(  R_{i}\cdot R_{h+n}\right)  \equiv
R_{i}\cdot\sum_{h=1}^{n}c_{h}R_{h+n}\equiv R_{i}\cdot Z_{a}\left(
\operatorname{mod}2\right)
\]
by Proposition \ref{invariant}. \ Therefore by checking whether $R_{i}%
$\ anticommutes with $Z_{a}$---which it does if and only if $x_{ia}=1$---we
learn whether $c_{i}=1$ and thus whether $\operatorname*{rowsum}\left(
2n+1,i+n\right)  $\ needs to be called.

We end this section by explaining how to compute the \textit{inner product}
between two stabilizer states $\left\vert \psi\right\rangle $\ and $\left\vert
\varphi\right\rangle $, given their full tableaus. \ The inner product is $0$
if the stabilizers contain the same Pauli operator with opposite signs.
\ Otherwise it equals $2^{-s/2}$, where $s$ is the minimum, over all sets of
generators $\left\{  G_{1},\ldots,G_{n}\right\}  $\ for\ $\operatorname*{Stab}%
\left(  \left\vert \psi\right\rangle \right)  $\ and $\left\{  H_{1}%
,\ldots,H_{n}\right\}  $\ for\ $\operatorname*{Stab}\left(  \left\vert
\varphi\right\rangle \right)  $, of the number of $i$ for which $G_{i}\neq
H_{i}$. \ For example, $\left\langle XX,ZZ\right\rangle $ and $\left\langle
ZI,IZ\right\rangle $ have inner product $1/\sqrt{2}$, since $\left\langle
ZI,IZ\right\rangle =\left\langle ZI,ZZ\right\rangle $. \ The proof is easy: it
suffices to observe that neither the inner product nor $s$ is affected if we
transform $\left\vert \psi\right\rangle $\ and $\left\vert \varphi
\right\rangle $\ to $U\left\vert \psi\right\rangle $\ and $U\left\vert
\varphi\right\rangle $ respectively, for some unitary $U$ such that
$U\left\vert \psi\right\rangle =\left\vert 0\right\rangle ^{\otimes n}$\ has
the trivial stabilizer. \ This same observation yields an algorithm to compute
the inner product: first transform the tableau of $\left\vert \psi
\right\rangle $ to that of $U\left\vert \psi\right\rangle =\left\vert
0\right\rangle ^{\otimes n}$\ using Theorem \ref{canonical}; then perform
Gaussian elimination on the tableau of $U\left\vert \varphi\right\rangle $ to
obtain $s$. \ Unfortunately, this algorithm takes order $n^{3}$\ steps.

\section{Implementation and Experiments\label{IMPLEM}}

We have implemented the tableau algorithm of Section \ref{SIM}\ in a C program
called CHP (CNOT-Hadamard-Phase), which is available for download \footnote{At
www.scottaaronson.com/chp}. \ CHP takes as input a
program in a simple \textquotedblleft quantum assembly
language,\textquotedblright\ consisting of four instructions: \texttt{c} $a$
$b$\ (apply CNOT from control $a$ to target $b$), \texttt{h} $a$\ (apply
Hadamard\ to $a$), \texttt{p} $a$\ (apply phase gate to $a$), and \texttt{m}
$a$\ (measure $a$ in the standard basis, output the result, and update the
state accordingly). \ Here $a$\ and $b$ are nonnegative integers indexing
qubits; the maximum $a$ or $b$ that occurs in any instruction is assumed to be
$n-1$, where $n$ is the number of qubits. \ As an example, the following
program demonstrates the famous quantum teleportation protocol of Bennett et
al. \cite{bbcjpw}:\bigskip

$\left.
\begin{tabular}
[c]{l}%
\texttt{h 1}\\
\texttt{c 1 2}%
\end{tabular}
\right\}  $
\begin{tabular}
[c]{l}%
EPR pair is prepared (qubit $1$ is\\
Alice's half; qubit $2$ is Bob's half)
\end{tabular}

$\left.
\begin{tabular}
[c]{l}%
\texttt{c 0 1}\\
\texttt{h 0}\\
\texttt{m 0}\\
\texttt{m 1}%
\end{tabular}
\right\}  $
\begin{tabular}
[c]{l}%
Alice interacts qubit $0$ (the state to\\
be teleported) with her half of the\\
EPR pair
\end{tabular}

$\left.
\begin{tabular}
[c]{l}%
\texttt{c 0 3}\\
\texttt{c 1 4}%
\end{tabular}
\right\}  $
\begin{tabular}
[c]{l}%
Alice sends $2$ classical bits to Bob
\end{tabular}

$\left.
\begin{tabular}
[c]{l}%
\texttt{c 4 2}\\
\texttt{h 2}\\
\texttt{c 3 2}\\
\texttt{h 2}%
\end{tabular}
\ \right\}  $
\begin{tabular}
[c]{l}%
Bob uses the bits from Alice to\\
recover the teleported state
\end{tabular}
\bigskip

We also have available CHP programs that demonstrate the Bennett-Wiesner dense
quantum coding protocol \cite{bw}, the GHZ (Greenberger-Horne-Zeilinger)
experiment \cite{ghz}, Simon's algorithm \cite{simon}, and the Shor $9$-qubit
quantum error-correcting code \cite{shor}.

Our main design goal for CHP was high performance with a large number of
qubits and frequent measurements. \ The only reason to use CHP instead of a
general-purpose quantum computer simulator such as QuIDD \cite{vmh}\ or QCL
\cite{oemer} is performance, so we wanted to leverage that advantage and make
thousands of qubits easily simulable rather than just hundreds. \ Also, the
results of Section \ref{PARITYL}\ suggest that classical postprocessing is
unavoidable for stabilizer circuits, since stabilizer circuits are not even
universal for classical computation. \ So if we want to simulate (for example)
Simon's algorithm, then one measurement is needed for each bit of the first
register. CHP's execution time will be dominated by these measurements, since
as discussed in Section \ref{SIM}, each unitary gate takes only $O\left(
n\right)  $\ time to simulate.

\begin{figure}
[ptb]
\begin{center}
\includegraphics[
trim=0.267406in 1.577672in 0.000000in 2.108410in,
height=2.8435in,
width=3.809in
]%
{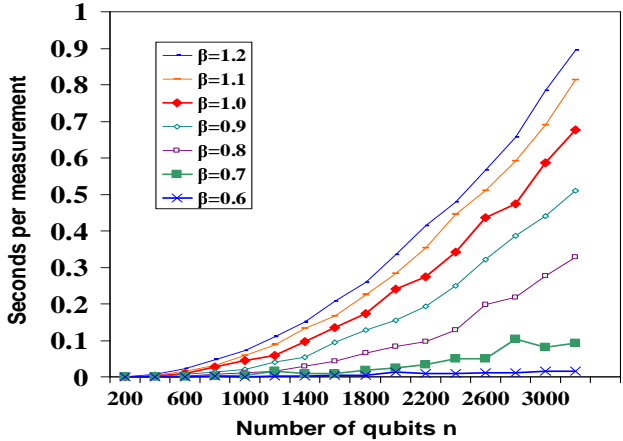}%
\caption{Average time needed to simulate a measurement after applying $\beta
n\log_{2}n$\ unitary gates to $n$ qubits,\ on a 650MHz Pentium III with 256MB
RAM.}%
\label{fig:chart}%
\end{center}
\end{figure}

Our experimental results, summarized in Figure \ref{fig:chart}, show that CHP
makes practical the simulation of arbitrary stabilizer circuits on up to about
$3000$ qubits. \ Since the number of bits needed to represent $n$ qubits grows
quadratically in $n$, the main limitation is available memory. \ On a machine
with 256MB of RAM, CHP can handle up to about $20000$\ qubits before virtual
memory is needed, in which case thrashing makes its performance intolerable.
\ The original version of CHP required \symbol{126}$8n^{2}$ bits for memory;
we were able to reduce this to \symbol{126}$4n^{2}$\ bits, enabling a 41\%
increase in the number of qubits for a fixed memory size. \ More trivially, we
obtained an eightfold improvement in memory by storing $8$ bits to each byte
instead of $1$. \ Not only did that change increase the number of storable
qubits by 183\%, but it also made CHP about 50\% faster---presumably because
(1) the $\operatorname*{rowsum}$\ subroutine now needed to exclusive-OR only
$1/8$\ as many bytes, and (2) the memory penalty was reduced. \ Storing the
bits in $32$-bit words yielded a further 10\% performance gain, presumably
because of (1) rather than (2) (since even with byte-addressing, a whole
memory line is loaded into the cache on a cache miss).

As expected, the experimentally measured execution time per unitary
gate\ grows linearly in $n$, whereas the time per measurement grows somewhere
between linearly and quadratically, depending on the states being measured.
\ Thus the time needed for measurements generally dominates execution time.
\ So the key question is this: what properties of a circuit determine whether
the time per measurement is linear, quadratic, or somewhere in between? \ To
investigate this question we performed the following experiment.

We randomly generated stabilizer circuits on $n$ qubits, for $n$ ranging from
$200$ to $3200$ in increments of $200$. \ For each $n$, we used the following
distribution over circuits: \textit{Fix a parameter }$\beta>0$\textit{; then
choose }$\left\lfloor \beta n\log_{2}n\right\rfloor $\textit{ random unitary
gates: a CNOT from control }$a$\textit{ to target }$b$\textit{, a Hadamard on
qubit }$a$\textit{, or a phase gate on qubit }$a$\textit{, each with
probability }$1/3$\textit{, where }$a$\textit{ and }$b$\textit{ are drawn
uniformly at random from }$\left\{  1,\ldots,n\right\}  $\textit{\ subject to
}$a\neq b$\textit{. \ Then measure qubit }$a$\textit{ for each }$a\in\left\{
1,\ldots,n\right\}  $\textit{ in sequence.}

We simulated the resulting circuits in CHP. \ For each circuit, we counted the
number of seconds needed for all $n$ measurement steps (ignoring the time for
unitary gates), then divided by $n$\ to obtain the number of seconds per
measurement. \ We repeated the whole procedure for $\beta$ ranging from
$0.6$\ to $1.2$\ in increments of $0.1$.

There were several reasons for placing measurements at the end of a circuit
rather than interspersing them with unitary gates. \ First, doing so models
how many quantum algorithms actually work (apply unitary gates, then measure,
then perform classical postprocessing); second, it allowed us to ignore the
effect of measurements on subsequent computation; third, it `standardized' the
measurement stage, making comparisons between different circuits more
meaningful; and fourth, it made simulation harder by increasing the propensity
for the measurements to be nontrivially correlated.

The decision to make the number of unitary gates\ proportional to $n\log
n$\ was based on the following heuristic argument. \ The time needed to
simulate a measurement is determined by how many times the
$\operatorname*{rowsum}$\ procedure is called, which in turn is determined by
how many $i$'s there are such that $x_{ia}=1$\ (where $a$ is the qubit being
measured). \ Initially $x_{ia}=1$\ if and only if $a=i$, so a measurement
takes $O\left(  n\right)  $\ time. \ For a random state, by contrast, the
expected number of $i$'s such that $x_{ia}=1$\ is $n$ by symmetry, so a
measurement takes order $n^{2}$\ time. \ In general, the more $1$'s there are
in the tableau, the longer measurements take. \ But where does the transition
from linear to quadratic time occur, and how sharp is it?

Consider $n$ people, each of whom initially knows one secret (with no two
people knowing the same secret). \ Each day, two people chosen uniformly at
random meet and exchange all the secrets they know. \ What is the expected
number of days until everyone knows everyone else's secrets? \ Intuitively,
the answer is $\Theta\left(  n\log n\right)  $, because any given person has
to wait $\Theta\left(  n\right)  $ days between meetings, and at each meeting,
the number of secrets he knows approximately doubles (or towards the end, the
number of secrets he \textit{doesn't} know is approximately halved).
\ Replacing people by qubits and meetings by CNOT gates, one can see why a
`phase transition' from a sparse to a dense tableau might occur after
$\Theta\left(  n\log n\right)  $\ random unitary gates are applied. \ However,
this argument does not pin down the proportionality constant $\beta$, so that
is what we varied in the experiment.

The results of the experiment are presented in Figure \ref{fig:chart}. \ When
$\beta=0.6$, the time per measurement appears to grow roughly linearly in $n$,
whereas when $\beta=1.2$ (meaning that the number of unitary gates has only
doubled), the time per measurement appears to grow roughly quadratically, so
that running the simulations took $4$ hours of computing time \footnote{Based
on our heuristic analysis, we conjecture that for intermediate $\beta$,\ the
time per measurement grows as $n^{c}$\ for some $1<c<2$. \ However, we do not
have enough data to confirm or refute this conjecture}. \ Thus, Figure
\ref{fig:chart} gives striking evidence for a \textquotedblleft phase
transition\textquotedblright\ in simulation time, as increasing the number of
unitary gates by only a constant factor shifts us from a regime of simple
states that are easy to measure, to a regime of complicated states that are
hard to measure. \ This result demonstrates that CHP's performance depends
strongly on the circuit being simulated. \ Without knowing what sort of
tableaus a circuit will produce, all we can say is that the time per
measurement will be somewhere between linear and quadratic in $n$.%

\section{Complexity of Simulating Stabilizer Circuits\label{PARITYL}}

The Gottesman-Knill theorem shows that stabilizer circuits are not universal
for quantum computation, unless quantum computers can be simulated efficiently
by classical ones. \ To a computer scientist, this theorem immediately raises
a question: where \textit{do} stabilizer circuits sit in the hierarchy of
computational complexity theory? \ In this section we resolve that question,
by proving that the problem of simulating stabilizer circuits is complete for
a classical complexity class known as $\mathsf{\oplus L}$\ (pronounced
\textquotedblleft parity-L\textquotedblright) \footnote{See
www.complexityzoo.com for definitions of $\mathsf{\oplus L}$\ and several
hundred other complexity classes}. \ The usual definition of $\mathsf{\oplus
L}$\ is as the class of all problems that are solvable by a nondeterministic
logarithmic-space Turing machine, that accepts if and only if the total number
of accepting paths is odd. \ But there is an alternate definition that is
probably more intuitive to non-computer-scientists. \ This is that
$\mathsf{\oplus L}$\ is the class of problems that reduce to simulating a
polynomial-size \textit{CNOT circuit}, i.e. a circuit composed entirely of NOT
and CNOT gates, acting on the initial state $\left\vert 0\cdots0\right\rangle
$. \ (It is easy to show that the two definitions\ are equivalent, but this
would require us first to explain what the usual definition \textit{means}!)

From the second definition, it is clear that $\mathsf{\oplus L}\subseteq
\mathsf{P}$; in other words, any problem reducible to simulating CNOT circuits
is also solvable in polynomial time on a classical computer. \ But this raises
a question: what do we mean by \textquotedblleft reducible\textquotedblright?
\ Problem $A$ is reducible to problem $B$ if any instance of problem $A$ can
be transformed into an instance of problem $B$; this means that problem $B$ is
\textquotedblleft harder\textquotedblright\ than problem $A$ in the sense that
the ability to answer an arbitrary instance of problem $B$ implies the ability
to answer an arbitrary instance of problem $A$ (but not necessarily vice-versa).

We must, however, insist that the reduction transforming instances of problem
$A$ into instances of problem $B$ not be too difficult to perform.
\ Otherwise, we could reduce hard problems to easy ones by doing all the
difficult work in the reduction itself. \ In the case of $\mathsf{\oplus L}$,
we \textit{cannot} mean \textquotedblleft reducible in polynomial
time,\textquotedblright\ which is a common restriction, since then the
reduction would be at least as powerful as the problem it reduces to!
\ Instead we require the reduction to be performed in the complexity class
$\mathsf{L}$, or \textit{logarithmic space}---that is, by a Turing machine $M$
that is given a read-only input of size $n$, and a write-only output tape, but
only $O\left(  \log n\right)  $\ bits of read/write memory. \ The reduction
works as follows: first $M$ specifies a CNOT circuit on its output tape; then
an \textquotedblleft oracle\textquotedblright\ tells $M$ the circuit's output
(which we can take to be, say, the value of the first qubit after the circuit
is applied), then $M$ specifies another CNOT circuit on its output tape, and
so on. \ A useful result of Hertrampf, Reith, and Vollmer \cite{hrv} says that
this seemingly powerful kind of reduction, in which $M$ can make multiple
calls to the CNOT oracle, is actually no more powerful than the kind with only
one oracle call. \ (In complexity language, what \cite{hrv}\ showed is that
$\mathsf{\oplus L}=\mathsf{L}^{\mathsf{\oplus L}}$: any problem in
$\mathsf{L}$\ with $\mathsf{\oplus L}$\ oracle is also in $\mathsf{\oplus L}$\ itself.)

It is conjectured that $\mathsf{L}\neq\mathsf{\oplus L}$; in other words, that
an oracle for simulating CNOT circuits would let an $\mathsf{L}$ machine
compute more functions than it could otherwise. \ Intuitively, this is because
writing down the intermediate states of such a circuit requires more than a
logarithmic number of read/write bits. \ Indeed, $\mathsf{\oplus L}$\ contains
some surprisingly \textquotedblleft hard\textquotedblright\ problems, such as
inverting matrices over $\mathbb{GF}_{2}$\ \cite{damm}. \ On the other hand,
it is also conjectured that $\mathsf{\oplus L}\neq\mathsf{P}$, meaning that
even with an oracle for simulating CNOT circuits, an $\mathsf{L}$\ machine
could not simulate more general circuits with AND and OR gates. \ As usual in
complexity theory, neither conjecture has been proved.

Now define the \textsc{Gottesman-Knill}\ problem as follows. \ We are given a
stabilizer circuit $\mathcal{C}$\ as a sequence of gates of the form
$\operatorname*{CNOT}$\ $a\rightarrow b$, $\operatorname*{Hadamard}$\ $a$,
$\operatorname*{Phase}$\ $a$, or $\operatorname*{Measure}$ $a$, where
$a,b\in\left\{  1,\ldots,n\right\}  $\ are indices of qubits. \ The problem is
to decide whether qubit $1$ will be $\left\vert 1\right\rangle $\ with
certainty after $\mathcal{C}$\ is applied to the initial state $\left\vert
0\right\rangle ^{\otimes n}$. \ (If not, then qubit $1$ will be $\left\vert
1\right\rangle $\ with probability either $1/2$ or $0$.)

Since stabilizer circuits\ are a generalization of CNOT circuits, it is
obvious that \textsc{Gottesman-Knill}\ is $\mathsf{\oplus L}$-hard\ (i.e. any
$\mathsf{\oplus L}$ problem can be reduced to it). \ Our result says that
\textsc{Gottesman-Knill}\ is \textit{in} $\mathsf{\oplus L}$. \ Intuitively,
this means that any stabilizer circuit can be simulated efficiently using CNOT
gates alone---the additional availability of Hadamard and phase gates gives
stabilizer circuits at most a polynomial advantage. \ In our view, this
surprising fact helps to explain the Gottesman-Knill theorem, by providing
strong evidence that stabilizer circuits are not even universal for
\textit{classical} computation (assuming, of course, that classical
postprocessing is forbidden).

\begin{theorem}
\label{parityl}\textsc{Gottesman-Knill} is in $\mathsf{\oplus L}$.
\end{theorem}

\begin{proof}
We will show how to solve \textsc{Gottesman-Knill}\ using a logarithmic-space
machine $M$ with an oracle for simulating CNOT circuits. \ By the result of
Hertrampf, Reith, and Vollmer \cite{hrv} described above, this will suffice to
prove the theorem.

By the principle of deferred measurement, we can assume that the stabilizer
circuit $\mathcal{C}$ has only a single measurement gate at the end (say of
qubit $1$), with all other measurements replaced by CNOT's into ancilla
qubits. \ In the tableau algorithm of Section \ref{SIM}, let $x_{ij}^{\left(
t\right)  },z_{ij}^{\left(  t\right)  },r_{i}^{\left(  t\right)  }$ be the
values of the variables $x_{ij},z_{ij},r_{i}$\ after $t$ gates of
$\mathcal{C}$\ have been applied. \ Then\ $M$ will simulate $\mathcal{C}$ by
computing these values. \ The first task of $M$ is to decide whether the
measurement has a determinate outcome---or equivalently, whether
$x_{i1}^{\left(  T\right)  }=0$\ for every $i\in\left\{  n+1,\ldots
,2n\right\}  $, where $T$ is the number of unitary gates. \ Observe that in
the CNOT, Hadamard, and phase procedures, every update to an $x_{ij}$\ or
$z_{ij}$\ variable replaces it by the sum modulo $2$ of one or two other
$x_{ij}$\ or $z_{ij}$\ variables. \ Also, iterating over all $t\in\left\{
0,\ldots,T\right\}  $ and $i\in\left\{  1,\ldots,2n\right\}  $\ takes only
$O\left(  \log n\right)  $\ bits of memory. \ Therefore, despite its memory
restriction, $M$\ can easily write on its output tape a description of a CNOT
circuit that simulates the tableau algorithm using $4n^{2}$ bits (the $r_{i}%
$'s being omitted), and that returns $x_{i1}^{\left(  T\right)  }$ for any
desired $i$. \ Then to decide whether the measurement outcome is determinate,
$M$ simply iterates over all $i$ from $n+1$\ to $2n$.

The hard part is to decide whether $\left\vert 0\right\rangle $\ or
$\left\vert 1\right\rangle $\ is measured in case the measurement outcome
\textit{is} determinate, for this problem involves the $r_{i}$\ variables,
which do not evolve in a linear way as the $x_{ij}$'s and $z_{ij}$'s do.
\ Even worse, it involves the complicated-looking and nonlinear
$\operatorname*{rowsum}$\ procedure. \ Fortunately, though, it turns out that
the measurement outcome $r_{2n+1}^{\left(  T+1\right)  }$ can be computed by
keeping track of a single complex number $\alpha$. \ This $\alpha$\ is a
product of phases of the form $\pm1$ or $\pm i$, and therefore takes only $2$
bits to specify. \ Furthermore, although the \textquotedblleft
obvious\textquotedblright\ ways to compute $\alpha$ use more than $O\left(
\log n\right)  $\ bits of memory, $M$ can get around that by making liberal
use of the oracle.

First $M$ computes what $r_{2n+1}^{\left(  T+1\right)  }$\ \textit{would}
be\ if the CNOT, Hadamard, and phase procedures did not modify the $r_{i}$'s.
\ Let $P$ be a Pauli matrix with a phase of $\pm1$ or $\pm i$, which therefore
takes $4$ bits to specify. \ Also, let $P_{ij}^{\left(  T\right)  }$\ be the
Pauli matrix represented by the bits $x_{ij}^{\left(  T\right)  }%
,z_{ij}^{\left(  T\right)  }$ in the usual way: $I=00$, $X=10$, $Y=11$,
$Z=01$. \ Then the procedure is as follows.\medskip

$\alpha:=1$

for $j:=1$\ to $n$

$\quad P:=I$

$\quad$for $i:=n+1$\ to $2n$\

$\quad\quad$ask oracle for $x_{\left(  i-n\right)  1}^{\left(  T\right)
},x_{ij}^{\left(  T\right)  },z_{ij}^{\left(  T\right)  }$

$\quad\quad$if $x_{\left(  i-n\right)  1}^{\left(  T\right)  }=1$ then
$P:=P_{ij}^{\left(  T\right)  }P$

$\quad$next $i$

$\quad$multiply $\alpha$\ by the phase of $P$ ($\pm1$ or $\pm i$)

next $j\medskip$

The \textquotedblleft answer\textquotedblright\ is $1$\ if $\alpha=-1$\ and
$0$\ if $\alpha=1$ (note that $\alpha$ will never be $\pm i$ at the end).
\ However, $M$ also needs to account for the $r_{i}$'s, as follows.\medskip

for $i:=n+1$\ to $2n$

$\quad$ask oracle for $x_{\left(  i-n\right)  1}^{\left(  T\right)  }$

$\quad$if $x_{\left(  i-n\right)  1}^{\left(  T\right)  }=1$

$\quad\quad$for $t:=0$\ to $T-1$

$\quad\quad\quad$if $\left(  t+1\right)  ^{st}$\ gate is a Hadamard or phase
on $a$

$\quad\quad\quad\quad$ask oracle for $x_{ia}^{\left(  t\right)  }%
,z_{ia}^{\left(  t\right)  }$

$\quad\quad\quad\quad$if $x_{ia}^{\left(  t\right)  }z_{ia}^{\left(  t\right)
}=1$\ then $\alpha:=-\alpha$

$\quad\quad\quad$end if

$\quad\quad\quad$if $\left(  t+1\right)  ^{st}$\ gate is a CNOT from $a$ to
$b$

$\quad\quad\quad\quad$ask oracle for $x_{ia}^{\left(  t\right)  }%
,z_{ia}^{\left(  t\right)  },x_{ib}^{\left(  t\right)  },z_{ib}^{\left(
t\right)  }$

$\quad\quad\quad\quad$if $x_{ia}^{\left(  t\right)  }z_{ib}^{\left(  t\right)
}\left(  x_{ib}^{\left(  t\right)  }\oplus z_{ia}^{\left(  t\right)  }%
\oplus1\right)  =1$\ then $\alpha:=-\alpha$

$\quad\quad\quad$end if

$\quad\quad$next $t$

$\quad$end if

next $i\medskip$

The measurement outcome, $r_{2n+1}^{\left(  T+1\right)  }$, is then $1$\ if
$\alpha=-1$\ and $0$\ if $\alpha=1$. \ As described above, the machine $M$
needs only $O\left(  \log n\right)  $\ bits to keep track of the loop indices
$i,j,t$, and $O\left(  1\right)  $\ additional bits to keep track of other
variables. \ Its correctness follows straightforwardly from the correctness of
the tableau algorithm.
\end{proof}

For a problem to be $\mathsf{\oplus L}$-complete simply means that it is
$\mathsf{\oplus L}$-hard\ \textit{and}\ in $\mathsf{\oplus L}$. \ Thus, a
corollary of Theorem \ref{parityl}\ is that \textsc{Gottesman-Knill} is
$\mathsf{\oplus L}$-complete.

\section{Canonical Form\label{CANONICAL}}

Having studied the simulation of stabilizer circuits, in this section we turn
our attention to \textit{manipulating} those circuits. \ This task is of
direct relevance to quantum computer architecture: because the effects of
decoherence build up over time, it is imperative (even more so than for
classical circuits) to minimize the number of gates as well as wires and other
resources. \ Even if fault-tolerant techniques will eventually be used to tame
decoherence, there remains the bootstrapping problem of building the
fault-tolerance hardware! \ In that regard we should point out that
fault-tolerance hardware is likely to consist mainly of CNOT, Hadamard, and
phase gates, since the known fault-tolerant constructions (for example, that
of Aharonov and Ben-Or \cite{ab}) are based on stabilizer codes. \

Although there has been some previous work on synthesizing CNOT circuits
\cite{iky,pmh,mn}\ and general classical reversible circuits \cite{spmh,lckl},
to our knowledge there has not been work on synthesizing stabilizer circuits.
\ In this section we prove a \textit{canonical form theorem} that is extremely
useful for stabilizer circuit synthesis. \ The theorem says that given
\textit{any} circuit consisting of CNOT, Hadamard, and phase gates, there
exists an equivalent circuit that applies a round of Hadamard gates only, then
a round of CNOT gates only, and so on in the sequence H-C-P-C-P-C-H-P-C-P-C.
\ One easy corollary of the theorem is that any tableau satisfying the
commutativity conditions of Proposition \ref{invariant} can be generated by
some stabilizer circuit. \ Another corollary is that any unitary stabilizer
circuit has an equivalent circuit with only $O\left(  n^{2}/\log n\right)  $\ gates.

Given two $n$-qubit unitary stabilizer circuits $\mathcal{C}_{1}%
,\mathcal{C}_{2}$, we say that $\mathcal{C}_{1}$ and $\mathcal{C}_{2}$\ are
\textit{equivalent}\ if $\mathcal{C}_{1}\left(  \left\vert \psi\right\rangle
\right)  =\mathcal{C}_{2}\left(  \left\vert \psi\right\rangle \right)  $\ for
all stabilizer states $\left\vert \psi\right\rangle $, where $\mathcal{C}%
_{i}\left(  \left\vert \psi\right\rangle \right)  $\ is the final state when
$\mathcal{C}_{i}$\ is applied to $\left\vert \psi\right\rangle $ \footnote{The
reason we restrict attention to unitary circuits is simply that, if
measurements are included, then it is unclear what it even \textit{means} for
two circuits to be equivalent. \ For example, does deferring all measurements
to the end of a computation preserve equivalence or not?}. \ By linearity, it
is easy to see that equivalent stabilizer circuits will behave identically on
\textit{all} states, not just stabilizer states. \ Furthermore, there exists a
one-to-one correspondence between circuits and tableaus:

\begin{lemma}
\label{equiv}Let $\mathcal{C}_{1},\mathcal{C}_{2}$\ be unitary stabilizer
circuits, and let $\mathcal{T}_{1},\mathcal{T}_{2}$\ be their respective final
tableaus when we run them on the standard initial tableau. \ Then
$\mathcal{C}_{1}$\ and $\mathcal{C}_{2}$\ are equivalent if and only if
$\mathcal{T}_{1}=\mathcal{T}_{2}$.
\end{lemma}

\begin{proof}
Clearly $\mathcal{T}_{1}=\mathcal{T}_{2}$\ if $\mathcal{C}_{1}$\ and
$\mathcal{C}_{2}$\ are equivalent. \ For the other direction, it suffices to
observe that a unitary stabilizer circuit acts linearly on Pauli operators
(that is, rows of the tableau): if it maps $P_{1}$\ to $Q_{1}$\ and $P_{2}%
$\ to $Q_{2}$,\ then it maps $P_{1}+P_{2}$\ to $Q_{1}+Q_{2}$. \ Since the rows
of the standard initial tableau form a basis for $\mathcal{P}_{n}$, the lemma follows.
\end{proof}

Our proof of the canonical form theorem will use the following two lemmas.

\begin{lemma}
\label{nonzero}Given an $n$-qubit stabilizer state, it is always possible to
apply Hadamard gates to a subset of the qubits so as to make the $X$ matrix
have full rank (or equivalently, make all $2^{n}$\ basis states have nonzero amplitude).
\end{lemma}

\begin{proof}
We can always perform row additions on the $n\times2n$\ stabilizer matrix
without changing the state that it represents. \ Suppose the $X$ matrix has
rank $k<n$; then by Gaussian elimination, we can put the stabilizer matrix in
the form%
\[
\left(
\begin{tabular}
[c]{l|l}%
$\,\,\,\,\,\,\,A\,\,\,\,\,\,\,$ & $\,\,\,\,\,\,\,B\,\,\,\,\,\,\,$\\
$\,\,\,\,\,\,\,0\,\,\,\,\,\,\,$ & $\,\,\,\,\,\,\,C\,\,\,\,\,\,\,$%
\end{tabular}
\right)
\]
where $A$ is $k\times n$\ and has rank $k$. \ Then since the rows are linearly
independent, $C$ must have rank $n-k$; therefore it has an $\left(
n-k\right)  \times\left(  n-k\right)  $\ submatrix $C_{2}$\ of full rank.
\ Let us permute the columns of the $X$ and $Z$ matrices simultaneously to
obtain%
\[
\left(
\begin{tabular}
[c]{cc|cc}%
$A_{1}$ & $A_{2}$ & $B_{1}$ & $B_{2}$\\
$0$ & $0$ & $C_{1}$ & $C_{2}$%
\end{tabular}
\right)  ,
\]
and then perform Gaussian elimination on the bottom $n-k$\ rows to obtain%
\[
\left(
\begin{tabular}
[c]{cc|cc}%
$A_{1}$ & $A_{2}$ & $B_{1}$ & $B_{2}$\\
$0$ & $0$ & $D$ & $I$%
\end{tabular}
\right)  .
\]
Now commutativity relations imply%
\[
\left(
\begin{array}
[c]{cc}%
A_{1} & A_{2}%
\end{array}
\right)  \left(
\begin{array}
[c]{c}%
D^{T}\\
I
\end{array}
\right)  =0
\]
and therefore $A_{1}D^{T}=A_{2}$. \ Notice that this implies that the $k\times
k$\ matrix $A_{1}$\ has full rank, since otherwise the $X$ matrix would have
column rank less than $k$. \ So performing Hadamards on the rightmost
$n-k$\ qubits yields a state%
\[
\left(
\begin{tabular}
[c]{cc|cc}%
$A_{1}$ & $B_{2}$ & $B_{1}$ & $A_{2}$\\
$0$ & $I$ & $D$ & $0$%
\end{tabular}
\right)
\]
whose $X$ matrix has full rank.
\end{proof}

\begin{lemma}
\label{gaussian}For any symmetric matrix $A\in\mathbb{Z}_{2}^{n\times n}$, there exists
a diagonal matrix $\Lambda$\ such that $A+\Lambda = M M^T$, with $M$ some
invertible binary matrix.
\end{lemma}

\begin{proof}
We will let $M$ be a lower-triangular matrix with $1$s all along the diagonal:
\begin{align}
M_{ii} = 1 & \\
M_{ij} = 0 & \quad \mbox{$i < j$}
\end{align}
Such an $M$ is always invertible. \ Then there exists a diagonal $\Lambda$ such that $A+\Lambda = M M^T$ if and only if
\begin{equation}
A_{ij} = \sum_{k} M_{ik} M_{jk}
\label{Meqs}
\end{equation}
for all pairs $(i,j)$ with $i > j$.  (We pick $\Lambda$ appropriately to satisfy the equations for $A_{ii}$ automatically, and both sides of the equation are symmetric, covering the cases with $i < j$.)

We will perform induction on $i$ and $j$ to solve for the undetermined elements of $M$.  For the base case, we know that $M_{11} = 1$.  We will determine $M_{ij}$ for $i > j$ by supposing we have already determined $M_{i'j'}$ for either $i' < i$, $j' \leq j$ or $i' \leq i$, $j' < j$.
We consider equation~(\ref{Meqs}) for $A_{ij}$ and note that $M_{ik} M_{jk} = 0$ unless $k \leq j$.  Then
\begin{equation}
A_{ij} = \sum_{k<j} M_{ik} M_{jk} + M_{ij}.
\end{equation}
By the induction hypothesis, we have already determined in the sum both $M_{ik}$ (since $k<j$) and $M_{jk}$ (since $j<i$ and $k<j$), so this equation uniquely determines $M_{ij}$.  We can thus find a unique $M$ that satisfies (\ref{Meqs}) for all $i > j$.
\end{proof}

Say a unitary stabilizer circuit is in \textit{canonical form} if it consists
of $11$ rounds in the sequence H-C-P-C-P-C-H-P-C-P-C.

\begin{theorem}
\label{canonical}Any unitary stabilizer circuit has an equivalent circuit in
canonical form.
\end{theorem}

\begin{proof}
Divide a $2n\times2n$\ tableau into four $n\times n$\ matrices $A=\left(
a_{ij}\right)  $, $B=\left(  b_{ij}\right)  $, $C=\left(  c_{ij}\right)  $,
and $D=\left(  d_{ij}\right)  $, containing the destabilizer $x_{ij}$ bits,
destabilizer $z_{ij}$\ bits, stabilizer $x_{ij}$\ bits, and stabilizer
$z_{ij}$\ bits\ respectively:%
\[
\left(
\begin{tabular}
[c]{l|l}%
$A$ & $B$\\\hline
$C$ & $D$%
\end{tabular}
\right)
\]
(We can ignore the phase bits $r_{i}$.) \ Since unitary circuits are
reversible, by Lemma \ref{equiv} it suffices to show how to obtain the
standard initial tableau starting from an arbitrary $A,B,C,D$ by applying CNOT, Hadamard, and phase gates
\footnote{Actually, this gives the canonical form for the inverse of the
circuit, but of course the same argument holds for the inverse circuit too,
which is also a stabilizer circuit}. \ We \textit{cannot} use row additions,
since although they leave states invariant they do not in general leave
circuits invariant.

The procedure is as follows.

\noindent\textbf{(1)} Use Hadamards to make $C$ have full rank (this is
possible by Lemma \ref{nonzero}).

\noindent\textbf{(2)} Use CNOT's to perform Gaussian elimination on $C$,
producing%
\[
\left(
\begin{tabular}
[c]{l|l}%
$A$ & $B$\\\hline
$I$ & $D$%
\end{tabular}
\right)  .
\]

\noindent\textbf{(3)} Commutativity of the stabilizer implies that
$ID^{T}$\ is symmetric, therefore $D$ is symmetric, and we can apply phase gates to add a diagonal matrix to $D$ and use
Lemma~\ref{gaussian} to convert $D$ to the form $D=MM^{T}$\ for
some invertible $M$.

\noindent\textbf{(4)} Use CNOT's to produce%
\[
\left(
\begin{tabular}
[c]{l|l}%
$A$ & $B$\\\hline
$M$ & $M$%
\end{tabular}
\right)  .
\]
Note that when we map $I$\ to $IM$, we also map $D$\ to $D\left(
M^{T}\right)  ^{-1}=MM^{T}\left(  M^{T}\right)  ^{-1}=M$.

\noindent\textbf{(5)} Apply phases to all $n$ qubits to obtain%
\[
\left(
\begin{tabular}
[c]{l|l}%
$A$ & $B$\\\hline
$M$ & $0$%
\end{tabular}
\right)  .
\]
Since $M$ is full rank, there exists some subset $S$\ of qubits such that
applying two phases in succession to every $a\in S$ will preserve the above
tableau, but set $r_{n+1}=\cdots=r_{2n}=0$. \ Apply two phases to every $a\in
S$.

\noindent\textbf{(6)} Use CNOT's to perform Gaussian elimination on $M$,
producing%
\[
\left(
\begin{tabular}
[c]{l|l}%
$A$ & $B$\\\hline
$I$ & $0$%
\end{tabular}
\right)  .
\]
By commutativity relations, $IB^{T}=A0^{T}+I$, therefore $B=I$.

\noindent\textbf{(7)} Use Hadamards to produce%
\[
\left(
\begin{tabular}
[c]{l|l}%
$I$ & $A$\\\hline
$0$ & $I$%
\end{tabular}
\right)  .
\]

\noindent\textbf{(8)} Now commutativity of the destabilizer implies that
$A$ is symmetric, therefore we can again use phase gates and Lemma~\ref{gaussian} to make $A=NN^{T}$\ for some invertible $N$.

\noindent\textbf{(9)} Use CNOT's to produce%
\[
\left(
\begin{tabular}
[c]{l|l}%
$N$ & $N$\\\hline
$0$ & $C$%
\end{tabular}
\right)  .
\]

\noindent\textbf{(10)} Use phases to produce%
\[
\left(
\begin{tabular}
[c]{l|l}%
$N$ & $0$\\\hline
$0$ & $C$%
\end{tabular}
\right)  ;
\]
then by commutativity relations, $NC^{T}=I$. \ Next apply two phases each to
some subset of qubits in order to preserve the above tableau, but set
$r_{1}=\cdots=r_{n}=0$.

\noindent\textbf{(11)} Use CNOT's to produce%
\[
\left(
\begin{tabular}
[c]{l|l}%
$I$ & $0$\\\hline
$0$ & $I$%
\end{tabular}
\right)  .
\]

\end{proof}

Since Theorem \ref{canonical}\ relied only on a tableau satisfying the
commutativity conditions, not on its being generated by some stabilizer
circuit, an immediate corollary is that any tableau satisfying the conditions
\textit{is} generated by some stabilizer circuit. \ We can also use Theorem
\ref{canonical} to answer the following question: how many gates are needed
for an $n$-qubit stabilizer circuit in the worst case? \ Cleve and Gottesman
\cite{cg} showed that $O\left(  n^{2}\right)  $\ gates suffice for the special
case of state preparation, and Gottesman \cite{gottesman3} and Dehaene and De
Moor \cite{dm}\ showed that $O\left(  n^{2}\right)  $\ gates suffice for
stabilizer circuits more generally; even these results were not obvious
\textit{a priori}. \ However, with the help of our canonical form theorem we
can show a stronger upper bound.

\begin{corollary}
\label{overlog}Any unitary stabilizer circuit has an equivalent circuit with
only $O\left(  n^{2}/\log n\right)  $\ gates.
\end{corollary}

\begin{proof}
Patel, Markov, and Hayes \cite{pmh}\ showed that any CNOT circuit has an
equivalent CNOT circuit with only $O\left(  n^{2}/\log n\right)  $\ gates.
\ So given a stabilizer circuit $\mathcal{C}$, first put $\mathcal{C}$ into
canonical form, then minimize the CNOT segments. \ Clearly the Hadamard and
Phase segments require only $O\left(  n\right)  $\ gates each.
\end{proof}

Corollary \ref{overlog}\ is easily seen to be optimal by a Shannon counting
argument: there are $2^{\Theta\left(  n^{2}\right)  }$\ distinct stabilizer
circuits on $n$ qubits, but at most $\left(  n^{2}\right)  ^{T}$ with $T$ gates.

A final remark: as noted by Moore and Nilsson \cite{mn}, any CNOT circuit has
an equivalent CNOT circuit with $O\left(  n^{2}\right)  $ gates and parallel
depth $O\left(  \log n\right)  $. \ Thus, using the same idea as in Corollary
\ref{overlog}, we obtain that any unitary stabilizer circuit has an equivalent
stabilizer circuit with $O\left(  n^{2}\right)  $\ gates and parallel depth
$O\left(  \log n\right)  $. \ (Moore and Nilsson showed this for the special
case of stabilizer circuits composed of CNOT and Hadamard gates only.)

\section{Beyond Stabilizer Circuits\label{BEYOND}}

In this section, we discuss generalizations of stabilizer circuits that are
still efficiently simulable. \ The first (easy) generalization, in Section
\ref{MIXED}, is to allow the quantum computer to be in a mixed rather than a pure
state. \ Mixed states could be simulated by simply purifying the state, and
then simulating the purification, but we present an alternative and slightly
more efficient strategy.

The second generalization, in Section \ref{INITIAL}, is to initial states
other than the computational basis state. \ Taken to an extreme, one could
even have noncomputable initial states. \ When combined with arbitrary quantum
circuits, such quantum advice is very powerful, although its exact power
(relative to classical advice) is unknown \cite{aaronson}. \ We consider a
more modest situation, in which the initial state may include specific ancilla
states, consisting of at most $b$ qubits each. \ The initial state is
therefore a tensor product of blocks of $b$ qubits. Given an initial state of
this form and general stabilizer circuits, including measurements and
classical feedback based on measurement outcomes, universal quantum
computation is again possible \cite{shor2,gc}. \ However, we show that an
efficient classical simulation exists, \textit{provided} only a few
measurements are allowed.

The final generalization, in Section \ref{GATES}, is to circuits containing a
few non-stabilizer gates. \ The qualifier \textquotedblleft
few\textquotedblright\ is essential here, since it is known that unitary
stabilizer circuits plus any additional gate yields a universal set of quantum
gates \cite{nrs,solovay}. \ The running time of our simulation procedure is
polynomial in $n$, the number of qubits, but is exponential in the $d$, the
number of non-stabilizer gates.

\subsection{Mixed States\label{MIXED}}

We first present the simulation for mixed states. \ We allow only
\textit{stabilizer mixed states}---that is, states that are uniform
distributions over all states in a subspace (or equivalently, all stabilizer
states in the subspace) with a given stabilizer of $r<n$ generators. \ Such
mixed states can always be written as the partial trace of a pure stabilizer
state, which immediately provides one way of simulating them.

It will be useful to see how to write the density matrix of the mixed state in
terms of the stabilizer. \ The operator $\left(  I+M\right)  /2$, when $M$ is
a Pauli operator, is a projection onto the $+1$ eigenspace of $M$.
\ Therefore, if the stabilizer of a pure state has generators $M_{1}%
,\ldots,M_{n}$, then the density matrix for that state is
\[
\rho=\frac{1}{2^{n}}\prod_{i=1}^{n}\left(  I+M_{i}\right)  .
\]
The density matrix for a stabilizer mixed state with stabilizer generated by
$M_{1},\ldots,M_{r}$ is
\[
\rho=\frac{1}{2^{r}}\prod_{i=1}^{r}\left(  I+M_{i}\right)  .
\]

To perform our simulation, we find a collection of $2\left(  n-r\right)  $
operators $\overline{X}_{i}$ and $\overline{Z}_{i}$ that commute with both the
stabilizer and the destabilizer. \ We can choose them so that $\left[
\overline{X}_{i},\overline{X}_{j}\right]  =\left[  \overline{Z}_{i}%
,\overline{Z}_{j}\right]  =\left[  \overline{X}_{i},\overline{Z}_{j}\right]
=0$ for $i\neq j$, but $\left\{  \overline{X}_{i},\overline{Z}_{i}\right\}
=0$. \ This can be done by solving a set of linear equations, which in
practice takes time $O\left(  n^{3}\right)  $. \ If we start with an initial
mixed state, we will assume it is of the form $\left\vert 00\cdots
0\right\rangle \left\langle 00\cdots0\right\vert \otimes I$ (so $0$ on the
first $n-r$ qubits and the completely mixed state on the last $r$ qubits).
\ In that case, we choose $\overline{X}_{i}=X_{i+r}$ and $\overline{Z}%
_{i}=Z_{i+r}$.

We could purify this state by adding $\overline{Z}_{i}Z_{n+i}$ and
$\overline{X}_{i}X_{n+i}$ to the stabilizer and $X_{n+i}$ and $Z_{n+i}$ to the
destabilizer for $i=1,\ldots,r$. \ Then we could simulate the system by just
simulating the evolution of this pure state through the circuit; the extra $r$
qubits are never altered.

A more economical simulation is possible, however, by just keeping track of
the original $r$-generator stabilizer and destabilizer, plus the $2\left(
n-r\right)  $ operators $\overline{X}_{i}$ and $\overline{Z}_{i}$. \ Formally,
this allows us to maintain a complete tableau and generalize the $O\left(
n^{2}\right)  $ tableau algorithm from Section~\ref{SIM}. \ We place the $r$
generators of the stabilizer as rows $n+1,\ldots,n+r$ of the tableau, and the
corresponding elements of the destabilizer as rows $1,\ldots,r$. \ The new
operators $\overline{X}_{i}$ and $\overline{Z}_{i}$ ($i=1,\ldots,n-r$) become
rows $r+i$ and $n+r+i$, respectively. \ Let $\overline{i}=i+n$ if $i\leq n$
and $\overline{i}=i-n$ if $i\geq n+1$. Then we have that rows $R_{i}$ and
$R_{j}$ commute unless $i=\overline{j}$, in which case $R_{i}$ and $R_{j}$ anticommute.

We can keep track of this new kind of tableau in much the same way as the old
kind. \ Unitary operations transform the new rows the same way as rows of the
stabilizer or destabilizer. \ For example, to perform a CNOT from control
qubit $a$ to target qubit $b$, set $x_{ib}:=x_{ib}\oplus x_{ia}$ and
$z_{ia}:=z_{ia}\oplus z_{ib}$, for all $i\in\left\{  1,\ldots,2n\right\}  $.

Measurement of qubit $a$ is slightly more complex than before. \ There are now
three cases:\medskip

\noindent\textbf{Case I:} $x_{pa}=1$ for some $p\in\left\{  n+1,\ldots
,n+r\right\}  $. \ In this case $Z_{a}$ anticommutes with an element of the
stabilizer, and the measurement outcome is random. \ We update as before, for
all rows of the tableau.

\noindent\textbf{Case II:} $x_{pa}=0\ $for all $p>r$. \ In this case$\ Z_{a}$
is in the stabilizer. \ The measurement outcome is determinate, and we can
predict the result as before, by calling $\operatorname*{rowsum}$ to add up
rows $r_{n+i}$ for those $i$ with $x_{ia}=1$.

\noindent\textbf{Case III: }$x_{pa}=0\ $for all $p\in\left\{  n+1,\ldots
,n+r\right\}  $, but $x_{ma}=1\ $for some $m\in\left\{  r+1,\ldots,n\right\}
$ or $m\in\left\{  n+r+1,\ldots,2n\right\}  $.\textbf{ } In this case $Z_{a}$
commutes with all elements of the stabilizer but is not itself in the
stabilizer. \ We get a random measurement result, but a slightly different
transformation of the stabilizer than in Case I. \ Observe that row $R_{m}$
anticommutes with $Z_{a}$. \ This row takes the role of row $p$ from Case I,
and the row $R_{\overline{m}}$ takes the role of row $p-n$. \ Update as before
with this modification. \ Then swap rows $n+r+1$ and $m$ and rows $r+1$ and
$\overline{m}$. \ Finally, increase $r$ to $r+1$: the stabilizer has gained a
new generator.\medskip

Another operation that we might want to apply is discarding the qubit $a$,
which has the effect of performing a partial trace over that qubit in the
density matrix. \ Again, this can be done by simply keeping the qubit in our
simulation and not using it in future operations. \ Here is an alternative:
put the stabilizer in a form such that there is at most one generator with an
$X$ on qubit $a$, and at most one with a $Z$ on qubit $a$. Then drop those two
generators (or one, if there is only one total). \ The remaining generators
describe the stabilizer of the reduced mixed state. \ We also must put the
$\overline{X}_{i}$ and $\overline{Z}_{i}$ operators in a form where they have
no entries in the discarded location, while preserving the structure of the
tableau (namely, the commutation relations of Proposition~\ref{invariant}).
\ This can also be done in time $O(n^{2})$, but we omit the details, as they
are rather involved.

\subsection{Non-Stabilizer Initial States\label{INITIAL}}

We now show how to simulate a stabilizer circuit where the initial state is
more general, involving non-stabilizer initial states. \ We allow any number
of ancillas in arbitrary states, but the overall ancilla state must be a
tensor product of blocks of at most $b$ qubits each. \ An arbitrary stabilizer
circuit is then applied to this state. \ We allow measurements, but only $d$
of them in total throughout the computation. \ We do allow classical
operations conditioned on the outcomes of measurements, so we also allow
polynomial-time classical computation during the circuit.

Let the initial state have density matrix $\rho$: a tensor product of $m$
blocks of at most $b$ qubits each. \ Without loss of generality, we first
apply the unitary stabilizer circuit $U_{1}$, followed by the measurement
$Z_{1}$ (that is, a measurement of the first qubit in the standard basis).
\ We then apply the stabilizer circuit $U_{2}$, followed by measurement
$Z_{2}$ on the second qubit, and so on up to $U_{d},Z_{d}$.

We can calculate the probability $p\left(  0\right)  $ of obtaining outcome
$0$ for the first measurement $Z_{1}$ as follows:
\begin{align*}
p\left(  0\right)   &  =\operatorname*{Tr}\left[  \left(  I+Z_{1}\right)
U_{1}\rho U_{1}^{\dagger}\right]  /2\\
&  =\operatorname*{Tr}\left[  \left(  I+U_{1}^{\dagger}Z_{1}U_{1}\right)
\rho\right]  /2\\
&  =1/2+\operatorname*{Tr}\left[  \left(  U_{1}^{\dagger}Z_{1}U_{1}\right)
\rho\right]  /2.
\end{align*}
But $U_{1}$ is a stabilizer operation, so $U_{1}^{\dagger}Z_{1}U_{1}$ is a
Pauli matrix, and is therefore a tensor product operation. \ We also know
$\rho$ is a tensor product of blocks of at most $b$ qubits, and the trace of a
tensor product is the product of the traces.\ \ Let $\rho=\otimes_{j=1}%
^{m}\rho_{j}$ and $U_{1}^{\dagger}Z_{1}U_{1}=\otimes_{j=1}^{m}P_{j}$ where $j$
ranges over the blocks. \ Then
\[
p\left(  0\right)  =\frac{1}{2}+\prod_{j=1}^{m}\operatorname*{Tr}\left(
P_{j}\rho_{j}\right)  .
\]
Since $P_{j}$ and $\rho_{j}$ are both $2^{b}\times2^{b}$-dimensional matrices,
each $\operatorname*{Tr}\left(  P_{j}\rho_{j}\right)  $ can be computed in
time $O\left(  2^{2b}\right)  $.

By flipping an appropriately biased coin, Alice can generate an outcome of the
first measurement according to the correct probabilities. \ Conditioned on
this outcome (say of $0$), the state of the system is
\[
\frac{\left(  I+Z_{1}\right)  U_{1}\rho U_{1}^{\dagger}\left(  1+Z_{1}\right)
}{4p\left(  0\right)  }\ .
\]
After the next stabilizer circuit $U_{2}$, the state is
\[
\frac{U_{2}\left(  I+Z_{1}\right)  U_{1}\rho U_{1}^{\dagger}\left(
1+Z_{1}\right)  U_{2}^{\dagger}}{4p\left(  0\right)  }\ .
\]
The probability of obtaining outcome $0$ for the second measurement,
conditioned on the outcome of the first measurement being $0$, is then
\[
p\left(  0|0\right)  =\frac{\operatorname*{Tr}\left[  \left(  I+Z_{2}\right)
U_{2}\left(  I+Z_{1}\right)  U_{1}\rho U_{1}^{\dagger}\left(  I+Z_{1}\right)
U_{2}^{\dagger}\right]  }{8p\left(  0\right)  }.
\]
By expanding out the $8$ terms, and then commuting $U_{1}$ and $U_{2}$ past
$Z_{1}$\ and $Z_{2}$, we can write this as%
\[
\sum_{i=1}^{8}\prod_{j=1}^{m}\operatorname*{Tr}\left(  P_{ij}^{\left(
2\right)  }\rho_{ij}\right)  .
\]
Each $\operatorname*{Tr}\left(  P_{ij}^{\left(  2\right)  }\rho_{ij}\right)  $
term can again be computed in time $O\left(  2^{2b}\right)  $.

Similarly, the probability of any particular sequence of measurement outcomes
$m_{1}m_{2}\cdots m_{d}$ can be written as a sum
\[
p\left(  m_{1}m_{2}\cdots m_{d}\right)  =\sum_{i=1}^{2^{2d-1}}\prod_{j=1}%
^{m}\operatorname*{Tr}\left(  P_{ij}^{\left(  d\right)  }\rho_{ij}\right)  ,
\]
where each trace can be computed in time $O\left(  2^{2b}\right)  $. \ It
follows that the probabilities of the two outcomes of the $d^{th}%
$\ measurement can be computed in time $O\left(  m2^{2b+2d}\right)  $.

\subsection{Non-Stabilizer Gates\label{GATES}}

The last case that we consider is that of a circuit containing $d$
non-stabilizer gates, each of which acts on at most $b$ qubits. \ We allow an
unlimited number of Pauli measurements and unitary stabilizer gates, but the
initial state is required to be a stabilizer state---for concreteness,
$\left\vert {0}\right\rangle ^{\otimes n}$.

To analyze this case, we examine the density matrix $\rho_{t}$ at the $t^{th}$
step of the computation. \ Initially, $\rho_{0}$ is a stabilizer state whose
stabilizer is generated by some $M_{1},\ldots,M_{n}$, so we can write it as
\[
\rho=\frac{1}{2^{n}}\left(  I+M_{1}\right)  \left(  I+M_{2}\right)
\cdots\left(  I+M_{n}\right)  .
\]
If we perform a stabilizer operation, the $M_{i}$'s become a different set of
Pauli operators, but keeping track of them requires at most $n\left(
2n+1\right)  $ bits at any given time (or $2n\left(  2n+1\right)  $ if we
include the destabilizer). \ If we perform a measurement, the $M_{i}$'s change
in a more complicated way, but remain Pauli group elements.

Now consider a single non-stabilizer gate $U$. Expanding $U$ in terms of Pauli
operations $P_{i}$,%
\begin{align*}
U\rho U^{\dagger}  &  =\frac{1}{2^{n}}\left(  \sum_{i}c_{i}P_{i}\right)
\prod_{j}\left(  I+M_{j}\right)  \left(  \sum_{k}c_{k}^{\ast}P_{k}\right) \\
&  =\frac{1}{2^{n}}\sum_{i,k}c_{i}c_{k}^{\ast}P_{i}P_{k}\prod_{j}\left(
I+\left(  -1\right)  ^{M_{j}\cdot P_{k}}M_{j}\right)  .
\end{align*}
Here $M_{j}\cdot P_{k}$ is the symplectic inner product between the
corresponding vectors, which is $0$ whenever $M_{j}$ and $P_{k}$ commute and
$1$ when they anticommute. \ In what follows, let $c_{ik}=c_{i}c_{k}^{\ast}%
$\ and $P_{ik}=P_{i}P_{k}$. \ Then we can write the density matrix after $U$
as a sum of terms, each described by a Pauli matrix $P_{ik}$ and a vector of
eigenvalues for the stabilizer. \ Since $U$ and $U^{\dagger}$ each act on at
most $b$ qubits, there are at most $4^{2b}$ terms in this sum.

If we apply a stabilizer gate to this state, all of the Pauli matrices in the
decomposition are transformed to other Pauli matrices, according to the usual
rules. \ If we perform another non-stabilizer gate, we can again expand it in
terms of Pauli matrices, and put it in the same form. \ The new gate can act
on $b$ new qubits, however, giving us more terms in the sum. \ After $d$ such
operations, we thus need to keep track of at most $4^{2bd}$ complex numbers
(the coefficients $c_{ik}$), $4^{bd}$ strings each of $2n$\ bits (the Pauli
matrices $P_{ik}$), and $4^{bd}$ strings each of $n$ bits\ (the inner products
$M_{j}\cdot P_{k}$). \ We also need to keep track of the stabilizer generators
$M_{1},\ldots,M_{n}$, and it will be helpful to also keep track of the
destabilizer, for a total of an additional $2n\left(  2n+1\right)  $ bits.

The above allows us to describe the evolution when there are no measurements.
\ What happens when we perform a measurement? \ Consider the unnormalized
density matrix corresponding to outcome $0$ for measurement of the Pauli
operator $Q$:
\[
\rho(0)=\frac{1}{2^{n+2}}Q^{+}\sum_{i,k}c_{ik}P_{ik}\prod_{j}\left(
I+(-1)^{M_{j}\cdot P_{k}}M_{j}\right)  Q^{+}%
\]
where here and throughout we let $Q^{+}=I+Q$ and $Q^{-}=I-Q$. \ As usual,
either $Q$ commutes with everything in the stabilizer, or $Q$ anticommutes
with some element of the stabilizer. \ (However, the measurement outcome can
be indeterminate in both cases, and may have a non-uniform distribution.) \ In
the first case, we can rewrite the density matrix as
\[
\rho\left(  0\right)  =\frac{1}{2^{n+2}}\sum_{i,k}c_{ik}Q^{+}P_{ik}Q^{+}%
\prod_{j}\left(  I+\left(  -1\right)  ^{M_{j}\cdot P_{k}}M_{j}\right)  .
\]
But $Q^{+}P_{ik}Q^{+}=2P_{ik}Q^{+}$ if $P_{ik}$ and $Q$ commute, and
$Q^{+}P_{ik}Q^{+}=Q^{+}Q^{-}P_{ik}=0$ if $P_{ik}$ and $Q$ anticommute.
\ Furthermore, as usual, as $Q$ commutes with everything in the stabilizer,
$Q$ is actually in the stabilizer, so projecting on $Q^{+}$ either is
redundant (if $Q$ has eigenvalue $+1$) or annihilates the state (if $Q$ has
eigenvalue $-1$). \ Therefore, we can see that $\rho\left(  0\right)  $ has
the same form as before:
\[
\rho\left(  0\right)  =\frac{1}{2^{n}}\sum_{i,k}c_{ik}P_{ik}\prod_{j}\left(
I+\left(  -1\right)  ^{M_{j}\cdot P_{k}}M_{j}\right)  ,
\]
where now the sum over $i$ is only over those $P_{ik}$'s that commute with
$Q$, and the sum over $k$ is only over those $P_{k}$'s that give eigenvalue
$+1$ for $Q$.

When $Q$ anticommutes with an element of the stabilizer, we can change our
choice of generators so that $Q$ commutes with all of the generators except
for $M_{1}$. \ Then we write $\rho\left(  0\right)  $ as:
\begin{align*}
\rho\left(  0\right)   &  =\frac{1}{2^{n+2}}\sum_{i,k}c_{ik}Q^{+}P_{ik}\left(
I+\left(  -1\right)  ^{M_{j}\cdot P_{k}}M_{1}\right)  Q^{+}\Lambda_{k}\\
&  =\frac{1}{2^{n+2}}\sum_{i,k}c_{ik}Q^{+}P_{ik}\left[  Q^{+}+\left(
-1\right)  ^{M_{j}\cdot P_{k}}Q^{-}M_{1}\right]  \Lambda_{k}%
\end{align*}
where%
\[
\Lambda_{k}=\prod_{j>1}\left(  I+\left(  -1\right)  ^{M_{j}\cdot P_{k}}%
M_{j}\right)  .
\]
If $P_{ik}$ and $Q$ commute, then we keep only the first term $Q^{+}$ in the
square brackets. If $P_{ik}$ and $Q$ anticommute, we keep only the second term
$Q^{-}M_{1}$ in the square brackets. \ In either case, we can rewrite the
density matrix in the same kind of decomposition:
\[
\rho\left(  0\right)  =\frac{1}{2^{n}}\sum_{i,k}c_{ik}P_{ik}Q^{+}\prod
_{j>1}\left(  I+\left(  -1\right)  ^{M_{j}\cdot P_{k}}M_{j}\right)  ,
\]
where $Q$ has replaced $M_{1}$ in the stabilizer, and any $P_{ik}$ that
anticommutes with $Q$ has been replaced by $P_{ik}M_{1}$, its corresponding
$c_{ik}$ replaced by $\left(  -1\right)  ^{M_{j}\cdot P_{k}}c_{ik}$.

Therefore, we can always write the density matrix after the measurement in the
same kind of sum decomposition as before, with no more terms than there were
before the measurement.\ \ The density matrices are unnormalized, so we need
to calculate $\operatorname*{Tr}\rho\left(  0\right)  $ to determine the
probability of obtaining outcome $0$. \ Computing the trace of a single term
is straightforward: it is $0$ if $P_{ik}$ is not in the stabilizer and
$\pm2^{n}c_{ik}$ if $P_{ik}$ is in the stabilizer (with $+$ or $-$ determined
by the eigenvalue of $P_{ik}$). \ To calculate $\operatorname*{Tr}\rho\left(
0\right)  $, we just need to sum the traces of the $4^{2bd}$ individual terms.
We then choose a random number to determine the actual outcome. \ Thereafter,
we only need to keep track of $\rho\left(  0\right)  $ or $\rho\left(
1\right)  $, which we can easily renormalize to have unit trace. \ Overall,
this simulation therefore takes time and space $O\left(  4^{2bd}%
n+n^{2}\right)  $.

\section{Open Problems\label{OPEN}}

\noindent\textbf{(1)} Iwama, Kambayashi, and Yamashita \cite{iky}\ gave a set
of \textit{local transformation rules} by which any CNOT circuit (that is, a
circuit consisting solely of CNOT gates) can be transformed into any
equivalent CNOT circuit. \ For example, a CNOT from $a$ to $b$ followed by
another CNOT from $a$ to $b$ can be replaced by the identity, and a CNOT from
$a$ to $b$ followed by a CNOT from $c$ to $d$ can be replaced by a CNOT from
$c$ to $d$ followed by a CNOT from $a$ to $b$, provided that $a\neq d$\ and
$b\neq c$. \ Using Theorem \ref{canonical}, can we similarly give a set of
local transformation rules by which any unitary stabilizer circuit can be
transformed into any equivalent unitary stabilizer circuit? \ Such a rule set
could form the basis of an efficient heuristic algorithm for minimizing
stabilizer circuits.

\noindent\textbf{(2)} Can the tableau algorithm be modified to compute
measurement outcomes in only $O\left(  n\right)  $\ time? \ (In case the
measurement yields a random outcome, updating the state might still take order
$n^{2}$\ time.)

\noindent\textbf{(3)} In Theorem \ref{canonical}, is the $11$-round sequence
H-C-P-C-P-C-H-P-C-P-C really necessary, or is there a canonical form that uses
fewer rounds? \ Note that if we are only concerned with state preparation, and
not with how a circuit behaves on any initial state other than the standard
one, then the $5$-round sequence H-P-C-P-H is sufficient.

\noindent\textbf{(4)} Is there a set of quantum gates that is neither
universal for quantum computation, \textit{nor} classically simulable in
polynomial time? \ Shi \cite{shi}\ has shown that if we generalize stabilizer
circuits by adding \textit{any} $1$- or $2$-qubit gate not generated by CNOT,
Hadamard, and phase, then we immediately obtain a universal set.

\noindent\textbf{(5)} What is the computational power of stabilizer circuits
with arbitrary tensor product initial states, but measurements delayed until
the end of the computation? \ It is known that, if we allow classical
postprocessing and control of future quantum operations conditioned on
measurement results, then universal quantum computation is possible
\cite{shor2,gc}. \ However, if all measurements are delayed until the end of
the computation, then\ the quantum part of such a circuit (though not the
classical postprocessing) can be compressed to constant depth. \ On the other
hand, Terhal and DiVincenzo \cite{td2} have given evidence that even
constant-depth quantum circuits might be difficult to simulate classically.

\noindent\textbf{(6)} Is there an efficient algorithm that, given a CNOT or
stabilizer circuit, produces an equivalent circuit of (approximately) minimum
size? \ Would the existence of such an algorithm have unlikely complexity
consequences? \ This might be related to the hard problem of proving
superlinear lower bounds on CNOT or stabilizer circuit size for explicit functions.

\section{Acknowledgments}

We thank John Kubiatowicz, Michael Nielsen, Isaac Chuang, Cris Moore, and
George Viamontes for helpful discussions, Andrew Cross for fixing an error
in the manuscript and software, and Martin Laforest for pointing out an error
in the proof of Theorem~\ref{canonical}. \ SA was supported by an NSF Graduate
Fellowship and by DARPA. \ DG is supported by funds from NSERC of Canada,\ and
by the CIAR in the Quantum Information Processing program.

\end{document}